\documentclass[conference]{IEEEtran}
\IEEEoverridecommandlockouts
\usepackage{amsmath,amssymb,amsfonts, amsthm}
\usepackage{graphicx}
\usepackage{booktabs}
\usepackage{makecell}
\usepackage{subfigure}
\usepackage{tabularx}

\graphicspath{{figs/}}

\begin{document}
\title{Connecting AI Learning and Blockchain Mining\\in 6G Systems}

\author{Yunkai~Wei, 
        Zixian~An, 
		Supeng~Leng, 
        and~Kun~Yang
\thanks{Yunkai Wei, Zixian An, and Supeng Leng are with School of Information and Communication Engineering, University of Electronic Science and Technology of China, Chengdu, China (e-mail: ykwei@uestc.edu.cn; zxan@std.uestc.edu.cn; spleng@uestc.edu.cn).}
\thanks{Kun Yang (corresponding author) is with School of Computer Science and Electronic Engineering, University of Essex, Colchester, CO4 3SQ, U.K. (e-mail: kunyang@essex.ac.uk).}
}

\markboth{Journal of \LaTeX\ Class Files,~Vol.~14, No.~8, August~2015}%
{Wei \MakeLowercase{\textit{et al.}}: Connecting AI Learning and Blockchain Mining in 6G Systems}

\maketitle

\begin{abstract}
The sixth generation (6G) systems are generally recognized to be established on ubiquitous Artificial Intelligence (AI) and distributed ledger such as blockchain.
However, the AI training demands tremendous computing resource, which is limited in most 6G devices. 
Meanwhile, miners in Proof-of-Work (PoW) based blockchains devote massive computing power to block mining, and are widely criticized for the waste of computation. 
To address this dilemma, we propose an Evolved-Proof-of-Work (E-PoW) consensus that can integrate the matrix computations, which are widely existed in AI training, into the process of brute-force searches in the block mining. 
Consequently, E-PoW can connect AI learning and block mining via the multiply used common computing resource. 
Experimental results show that E-PoW can salvage by up to 80 percent computing power from pure block mining for parallel AI training  in 6G systems.
\end{abstract}


\IEEEpeerreviewmaketitle

\section{Introduction}
\IEEEPARstart{T}{he} sixth generation (6G) systems are generally recognized to be established on ubiquitous Artificial Intelligence (AI) to achieve efficient networking, communicating and data analyzing \cite{CM_2019_AI_empowered_6G}, as well as on distributed ledger such as blockchain to ensure the security, throughput, reliability, trust, and transparency of the systems \cite{TNSE_2020_BATS}. 
However, the training of AI demands huge computing resource, which is usually limited in most 6G devices, and should be assisted by cloud computing, fog computing or edge computing. This will eventually raise the cost of 6G systems in many aspects, such as the construction, operation, and utilization. 
Meanwhile, in Proof-of-Work (PoW) \cite{bitcoin} based blockchains, each miner is equipped with generous computing resource to execute a brute-force search for the target hash value, and thereby competes to generate a valid block. 
Although widely adopted for better decentralization and security over alternative consensuses \cite{coincap} such as Proof-of-Stake (PoS) \cite{ppcoin} or Proof-of-Activity (PoA) \cite{poa}, PoW is frequently criticized for tremendous waste in the computing resource. Some novel consensus with identical performance but much less computation consumption than PoW is urgent to be developed.

Studies have been investigated to reduce such resource waste in blockchains and save more computing power for AI training. Lightweight consensuses such as PoS \cite{ppcoin}, PoA \cite{poa}, Delegated-PoS \cite{dpos}, and Proof-of-Luck \cite{PoL}, are proposed to replace PoW. A profit optimizing game between block mining and AI training service providing is studied for the miners in \cite{Wei_2020_Profit_game}. 
Whereas, these consensuses may considerably degrade the decentralization or security of the blockchain system.
Instead of decreasing the computing resource consumed by the consensuses, few studies try to salvage the wasted computing power in PoW with valuable task processing.
In \cite{pox}, a Proof-of-Exercise (PoX) consensus uses a pool of task proposals to replace the target hash value search by computing tasks, which however have different complexity levels and lead to unfair competition \cite{consensus_survey}.
Primecoin \cite{primecoin} replaces PoW's target hash value search with the search for two special chains of prime number. 
The Proof-of-Deep-Learning (PoDL) consensus \cite{podl} requires the miners to participate in a competition of deep learning model training to generate valid blocks. 
However, the amount of computing resource salvaged by either Primecoin or PoDL is severely limited, since all the miners are solving the identical problem simultaneously to keep the fairness of the block generating competition.

\begin{figure*}
	\center
	\includegraphics[width=0.88\textwidth]{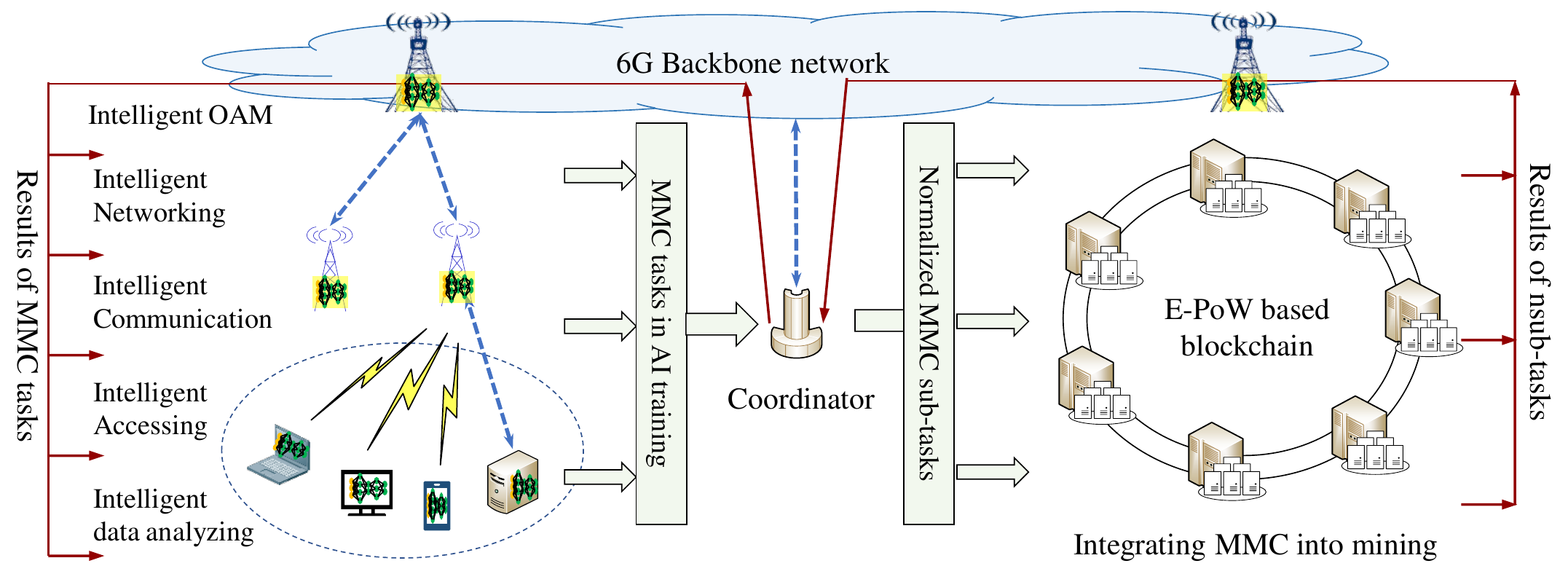}
	\caption{The basic idea of E-PoW to connect AI training and block mining in 6G systems.}
	\label{Fig::E-PoW-arch}
\end{figure*}

To address this dilemma in both AI and blockchain, we propose an Evolved-Proof-of-Work (E-PoW) consensus to connect the computing  resource consumed by AI training and block mining, and thereby improve the computing efficiency in 6G systems.
The underlying connection is the matrix multiplication calculation (MMC). MMC exists widely in AI. As an example, nearly 90\% of the workload in Google's Tensor Processing Unit is due to multi-layer perceptrons (MLP) and recurrent neural networks (RNN), which are both deep learning algorithms based on MMC \cite{google}. 
At the same time, MMC is feasible to be quantified and integrated into the process of the brute-forth search for the target hash value.
Consequently, E-PoW can integrate the massive MMC of AI training into the block mining of PoW-based blockchains, keeping the advantages of PoW as well as making the computing power efficiently utilized. The contributions of this paper are summarized as below:

\begin{itemize}
	\item We propose a novel consensus E-PoW, where MMC in AI training is integrated into the block mining process, and the miners conduct the target hash value search based on both the traditional block header and the result of MMC.
	\item We design detailed schemes to transform MMC tasks with different computing complexities into normalized matrix multiplications, which have identical computing complexity and can be easily integrated into the process of block mining through a uniform interface. 
	\item We conduct experiments in a campus network to verify the availability and effect of E-PoW. The results show that E-PoW can salvage considerable computing power from pure block mining for parallel AI training.
\end{itemize}

The remainder of this paper is organized as follows. 
The essentials of E-PoW are demonstrated in Section II.
In section III, we design detailed schemes to normalize and integrate MMC into E-PoW.
Section IV shows the performance of E-PoW with experiments. 
Finally, section V concludes this paper.

\section{Essentials of E-PoW}
In this section, we introduce the essentials of E-PoW, which can efficiently integrate the MMC tasks generated by AI training into the brute-force hash search in blockchain. 

\subsection{Architecture}
As shown in Fig.~\ref{Fig::E-PoW-arch}, AI training for intelligent processes in 6G systems, such as intelligent Operation, Administration and Maintenance (OAM), networking, communicating, accessing, and data analyzing, continuously generates massive MMC requirements, which are injected as computing tasks into an E-PoW based blockchain. An intermediate node, or so called the coordinator, is responsible for dividing these tasks into normalized sub-tasks (nsub-tasks), distributing these nsub-tasks to the miners, integrating the results submitted by the miners, and returning the final results to the task source in 6G systems. 
During every target hash value search period, each miner calculates the nsub-task it claims, returns the result to the coordinator for task computing reward, and conducts the brute-force search based on traditional PoW header, nsub-task information, and the hash of this nsub-task's result.

\subsection{Coordinator}
The coordinator divides and normalizes the MMC tasks into nsub-tasks, and distributes these nsub-tasks to the miners. 
In addition, it needs to verify (in a way with low overhead) and merge the results from the miners before returning them to the MMC task sources in the 6G system. For possible inquiries and verifications, the coordinator also stores the information of nsub-tasks and their results for feasible verification periods.

Specifically, an MMC task is generally the product of more than two matrices. The coordinator first divides it into a series of sub-tasks, each being the product of two matrices. These sub-tasks, with different sizes and computing complexities, are further divided into nsub-tasks, which have uniform size and computing complexity. 
Identified with (task id, sub-task id, nsub-task id), these nsub-tasks are sequentially calculated by the miners, who will subsequently return back the results. 
Then, the coordinator verifies and merges these results into the final result for the original MMC task. 
A verified result can bring corresponding reward to the miner calculating it, while a fake result would make this miner get penalty instead of reward for this nsub-task. 

\subsection{Miners}
An E-PoW enabled miner may conduct two types of loops to execute a brute-force search for the target hash value, namely task-free loop and task-involved loop.
In the task-free loop, the miner does not participate in MMC tasks and acts as a traditional PoW miner.
In the task-involved loop, the miner executes the brute-force search based not only on the traditional PoW's header (including the nonce), but also on a Learning-Service-Providing (LSP) field consisting of $5$ elements: task ID, sub-task ID, nsub-task ID, hash of nsub-task, hash of nsub-task result.
To keep the fairness of the miners in generating valid blocks, a difficulty re-targeting scheme is adopted to set feasible hash difficulties for the task-involved loops.

In the beginning of a block time for generating a valid block, the miner first decides the number of the nsub-tasks it intends to complete during this block time. Then, the miner will conduct the corresponding number of task-involved brute-force loops, each loop including one nsub-task, with one of the two possible cases following:
\begin{itemize}
\item The new valid block is successfully generated before the task-involved loops are completed. The miner will broadcast (or verify, when this miner is not the generator) this block, extend the left nsub-tasks to the next block time, and decrease the number of nsub-tasks it claims for the next block time.
\item The task-involved loops are completed while the new valid block is not yet successfully generated. The miner will conduct task-free loops until the new block is broadcast by itself or some other miner.
\end{itemize}

In each task-involved loop, the miner completes the nsub-task of this loop it fetches from the coordinator, fills in the LSP field based on the information and result matrix of this nsub-task, and submits the later to the coordinator. If the result matrix is verified, the miner will get the corresponding task reward. Otherwise, the nsub-tasks of the miner in this block time will be shifted to other available miners and no new nsub-tasks will be assigned to this miner in the next block time. The validation of the block generated in a task-involved loop should include the LSP field, in which the task information can be inquired from the coordinator.

\subsection{Decentralization and Security of E-PoW}
Besides the capability of connecting AI and blockchain in 6G systems, E-PoW can simultaneously keep the advantages of PoW in aspects of both decentralization and security, which are two major considerations in blockchain systems.

\subsubsection{Decentralization}
In an E-PoW based blockchain, each miner has proportional possibility corresponding to its computing power in generating a valid block, based on the follows:
\begin{itemize}
\item Ensured fairness. Each task-free loop has identical computing complexity with any others, keeping the fairness among them. And so do the task-involved loops in one miner and one block time, since the nsub-task involved in each task-involved loop also has identical computing complexity. For the fairness between task-free loops and task-involved loops, as well as the fairness among task-involved loops in different miners and different block times, a hash difficulty re-targeting scheme is proposed and will be described in detail in Section III.C.
\item Independent validation. Although an intermediate node named coordinator is introduced for task assigning and result fetching, the independence of new block validation will not be influenced. From the validating nodes' perspective, the data and result of one task can be identified with the Hash scheme.
\item Honest coordinator. The coordinator is honest since:  \textcircled{1} it can be a server from a trustful third party; \textcircled{2} although not designed in detail, the honesty can be ensured based on a simple surveillance scheme by the MMC task sources; \textcircled{3} the coordinator needs the system operating smoothly to achieve more profit from the reward difference between the MMC task sources and the miners.  
\end{itemize}

\subsubsection{Security}
In PoW-based blockchains, 51 percent attack is the major security threat. 
The schemes in E-PoW have not decreased the decentralization of PoW, and an E-PoW based blockchain's tolerated power of the adversary is also no less than 51 percent. 
In addition, E-PoW based blockchain is an endogenously increasing system since its mutually beneficial scheme can continuously enroll new positive participants pursuing the conveniences or the profits. The additional revenue of completing an MMC task is deterministic and friendly to the miners with limited computing power, which, in PoW-based blockchains, would fall through for extremely low revenue from block mining and thereby had more impulsion to take adversarial behaviors.
All these factors can cooperatively increase the security of E-PoW based blockchains.

\section{Process of Matrix Multiplication Calculation}
MMC is the underlying connection for E-PoW to integrate the AI training into the block mining. In this section, we will present the detailed MMC processing schemes in E-PoW.

\subsection{Normalization of MMC Tasks}
An original MMC task that calculates the multiplication of more than $2$ matrices can be divided into a series of sub-tasks by the coordinator, each sub-task being the multiplication of $2$ matrices.
The first sub-task is calculating the multiplication of the first two matrices in the original task, the second sub-task is the multiplication of the first sub-task's result matrix and the third matrix in the original task, and so on.
Consequently, the last sub-task, whose result equals to that of the original task, multiplies the result matrix of its previous sub-task by the last matrix of the original task.

Each sub-task will be further divided into a series of normalized nsub-tasks. Each nsub-task is the multiplication of two square matrices with identical number of rows and columns (or basic size for brevity). Consequently, the nsub-tasks have uniform size and computing complexity. 
The basic size can be determined according to the application environment. 
For example, in a scenario that the numbers of rows and columns in the sub-tasks' matrices vary from several thousand to ten thousand, the basic size can be set to $500$. The scheme of dividing a sub-task into nsub-tasks contains two steps: matrix expansion and matrix partition.

\begin{figure}
	\centerline{\includegraphics[width=0.48\textwidth]{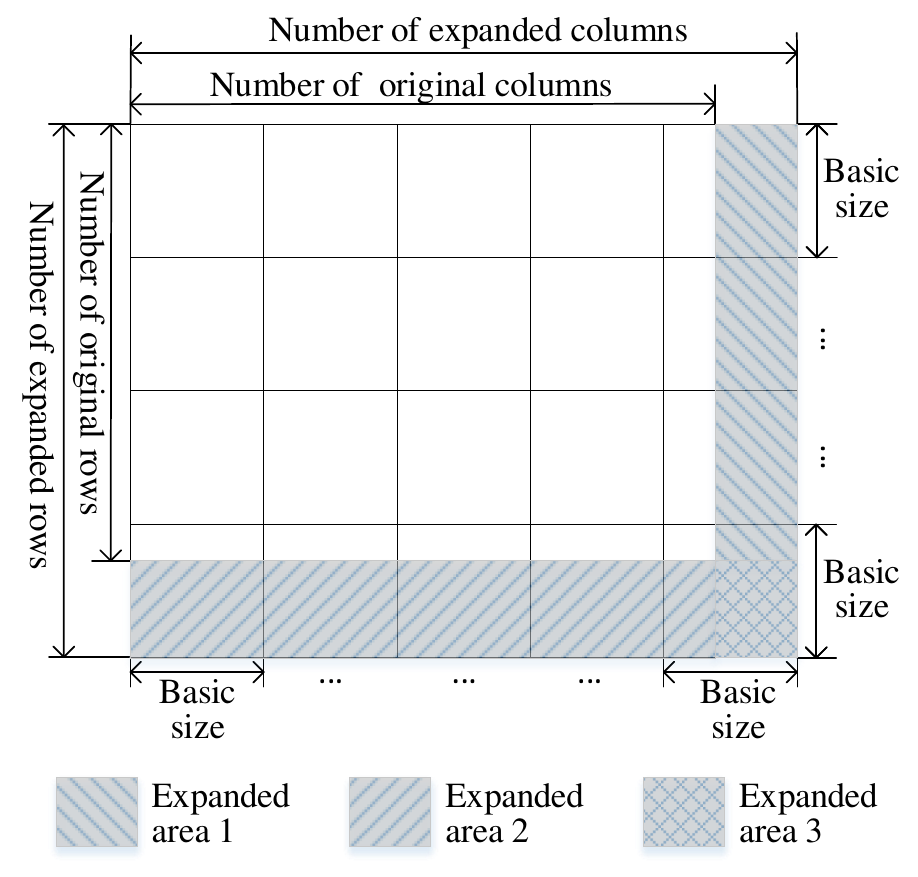}}
	\caption{Expansion and partition of the matrix in a sub-task.}
	\label{Fig::matrix-norm}
\end{figure}

\subsubsection{Matrix expansion}
As depicted above, each sub-task is the multiplication of two matrices. Each matrix will be firstly expanded in a minimum necessary extent, so that its numbers of expanded rows and columns are both the integral multiples of the basic size, as shown in Fig.~\ref{Fig::matrix-norm}. The expanded areas can be randomly generated, as long as the following condition holds: the multiplication of the expanded area $1$ in the first matrix of the sub-task and the expanded area $2$ in the second matrix of the sub-task is a zero matrix. In this way, the sub-task's result matrix is exactly the sub-matrix in the upper left corner of the multiplication of the two expanded matrices.

\subsubsection{Matrix partition}
The two expanded matrices of one sub-task, whose numbers of rows and columns are all integral multiples of the basic size, can be partitioned without overlap into basic square matrices with the basic size, as shown in Fig.~\ref{Fig::matrix-norm}.
According to the multiplication principles of matrices, the result of the sub-task can be obtained by calculating the corresponding multiplications of these basic square matrices.
That is to say, the sub-task can be divided into appropriate nsub-tasks, each nsub-task calculating the multiplication of two basic square matrices with uniform basic size, and thereby these nsub-tasks have identical complexity and workload.
By properly merging these nsub-tasks' result matrices, we can easily get the result matrix of the sub-task.

\subsection{Result Verification}
The normalized nsub-tasks are continuously assigned by the coordinator to the miners according to each miner's initiative on finishing MMC tasks. In order to ensure that the miners complete the nsub-tasks honestly, a result verification scheme is needed to check the correctness of the result matrices submitted by the miners.
The verification of the result for a nsub-task may be conducted by the coordinator or other miners. In some extreme environment with adversaries aiming at destroying the AI training instead of cheating for the reward, existing verification algorithms such as the Freivalds’ algorithm, can be adopted. Otherwise, we can use a verification scheme with negligible computing overload as follows. 

The nsub-task to be verified is calculating the multiplication of two square matrices with basic size, and the result matrix of this nsub-task is also a square matrix with basic size. 
Randomly choose a location in the result matrix, calculate the element value at this location based on the original square matrices in the nsub-task, and compare this verification value with the element in the result matrix submitted by the miner. 
If the verification value doesn't equal to the corresponding element in the result matrix, the result matrix is wrong, and the miner is suspected of cheating for the reward. 
The randomness in the location choosing, and several times repeating of the verification as needed, can ensure the accuracy of the verification process. 

\subsection{Difficulty Re-Targeting}
In an E-PoW based blockchain system, the difficulty for the miners to generate a new block can decide the block generation rate (BGR). It is influenced by the basic difficulty and a scale factor, as follows.

The system adjusts the basic difficulty every time when a given number of valid blocks have been successfully generated. 
This number is called the basic difficulty re-targeting (BDRT) window. 
The basic difficulty in current BDRT window is determined by three elements: the basic difficulty in the previous BDRT window, the target BGR in the previous BDRT window, and the actual BGR in the previous BDRT window. 
If the system's actual BGR is larger than the target BGR, which means that the system generates blocks faster than expected, the basic difficulty will be increased, and vice versa.

The scale factor for a miner in one block time is a value varies from $0$ to $1$, and is inversely proportional to the number of nsub-tasks the miner wants to complete in this block time.
Consequently, a miner with lower enthusiasm in completing nsub-tasks will have a larger scale factor. When the miner chooses not to participate in completing nsub-tasks, the scale factor is the maximum value $1$.

Then, the final difficulty for a miner to generate a new valid block equals to the multiplication of the basic difficulty and the scale factor.
The basic difficulty is same for all the miners, but the scale factor will change with the number of nsub-tasks that the miners want to complete.
Since the scale factor is inversely proportional to this number, as a consequence, the final difficulty is inversely proportional to this number, which means that the more active the miners are involved in completing nsub-tasks, the lower their mining difficulty will be.
Furthermore, when all the miners won't take any nsub-tasks, their scale factors will be $1$, and the E-PoW based system will degenerate into a traditional PoW-based system, where all the miners have the same hash difficulty, which equals to the basic difficulty. 

\section{Experiment and Results Analyses}
In this section, we conduct experiments to evaluate the availability and performance of the E-PoW consensus. 
We will firstly introduce the experimental environment, and then demonstrate the experiment results and analyses in detail.

\subsection{Experimental Environment}
\begin{figure}
	\centerline{\includegraphics[width=0.48\textwidth]{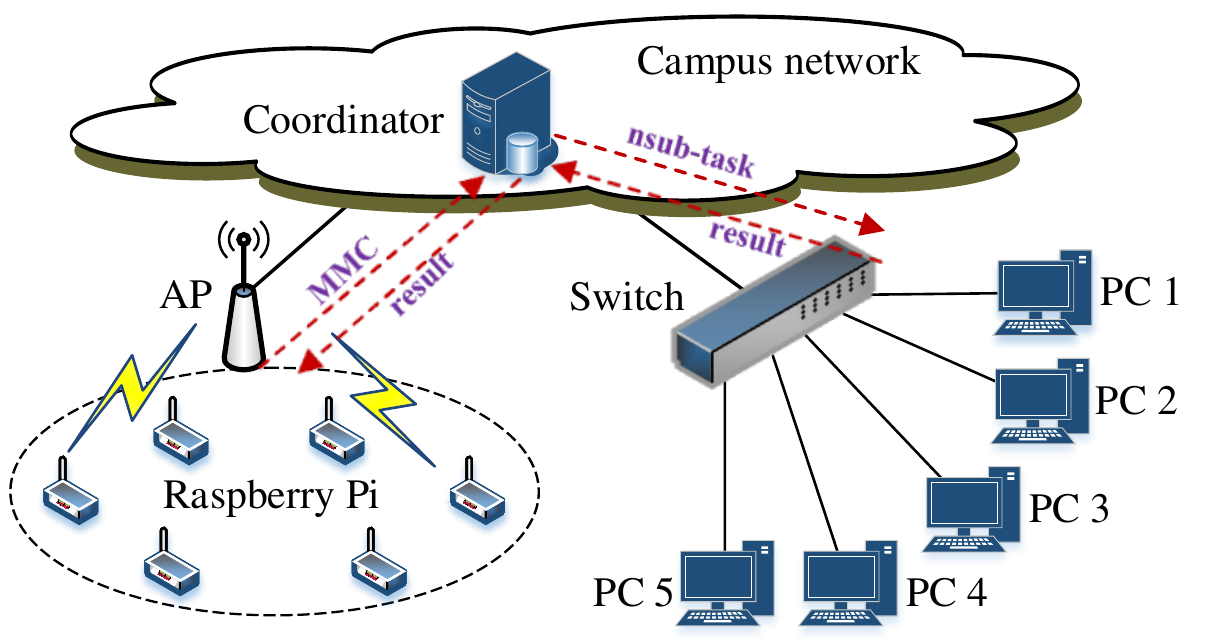}}
	\caption{The experimental environment.}
	\label{Fig::Exp_environ}
\end{figure}

We build the experimental platform in a campus network. 
As shown in Fig.~\ref{Fig::Exp_environ}, six Raspberry Pi Model 3B act as user devices running AI training and processing, which can continuously generate MMC tasks. A ThinkSystem SR860 server acts as the coordinator, transforming the MMC tasks into nsub-tasks for the miners and returning the results back to the user devices. Five personal computers with Intel Core i5-6500 CPU @ 3.20GHz, which are connected through an Ethernet switch, play the part of five miners in an E-PoW based blockchain.

In our experiment, the square matrices are set to have $500$ rows and $500$ columns, and the blockchain's initial mining difficulty is set to $ 2.5 \times 10^{-3} $.
According to the hardware of our experimental platform, it takes about 0.48 second for the miners to execute one task-involved loop, and the target average BGR is set to $ 1/300 $ block per second. Consequently, the maximum number of nsub-tasks that each miner can complete in a block time is set to $ 600 $.
We run the E-PoW consensus in five types of blockchain networks. The first four types are Type I, II, III and IV network, composed of Type I, II, III, IV miners, respectively. 
The difference among the four kinds of miners is the initiative of completing the nsub-tasks. For a Type I miner, in each block time it claims $0$ nsub-task. And for Type II, III and IV miners, the figures are $200$, $400$ and $600$, respectively.
The last one is Type V network, composed of all the above four kinds of miners. The number of miners in the blockchain network varies in a range from 2 to 5 to show the performance changing of E-PoW when the network scale is changing.



\subsection{Experiment Results and Analyses}
We conduct the experiment for $50$ times, and show the average results and the analyses in two aspects, namely the time of generating a block, and the efficiency of the miners' computing power.
\subsubsection{Time of generating a block}
\begin{figure}
	\centerline{\includegraphics[width=0.48\textwidth]{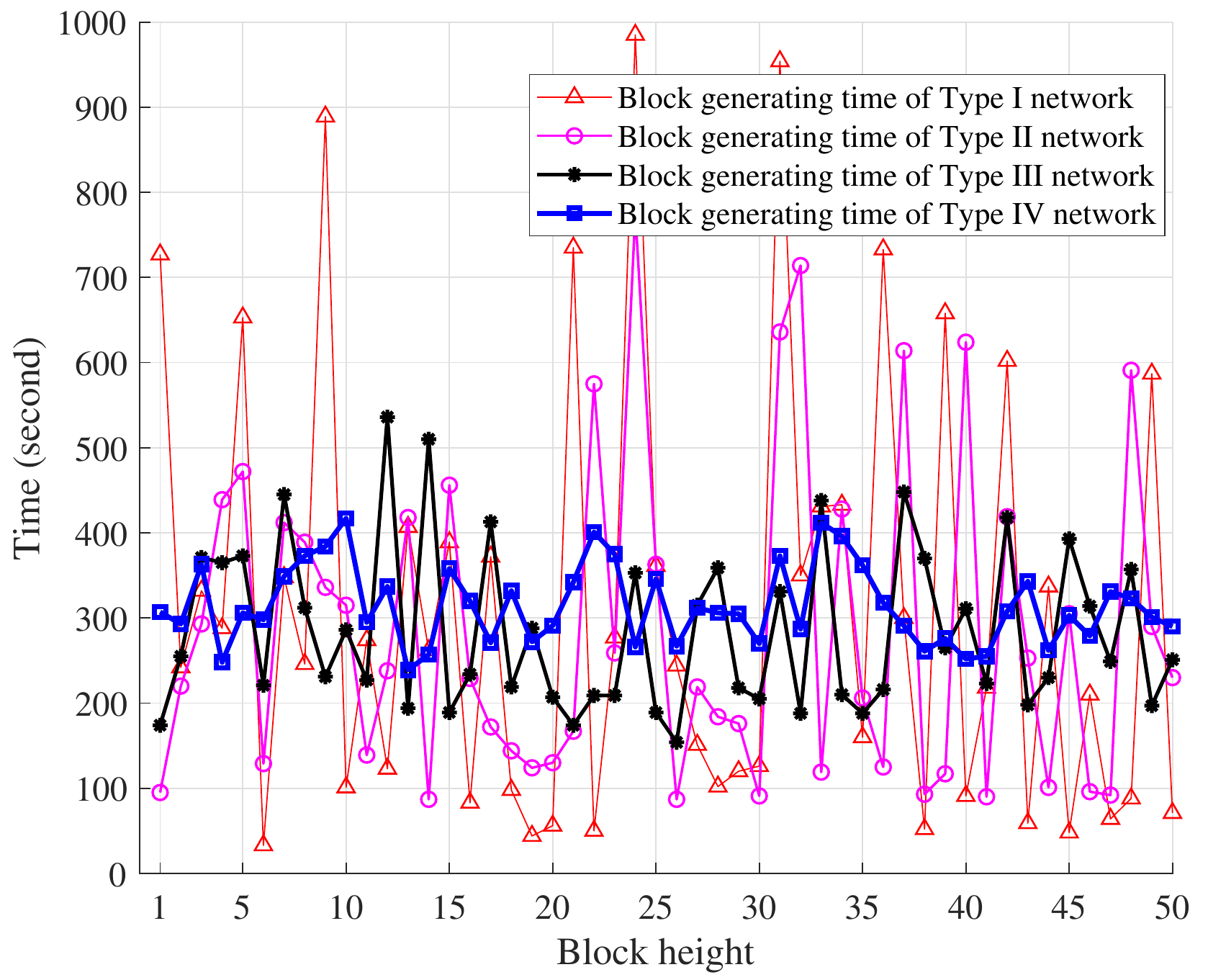}}
	\caption{The block height versus block generating time.}
	\label{BlockTime}
\end{figure}

Fig.~\ref{BlockTime} shows the time to generate a valid block in Type I, II, III and IV network respectively, where each type of network contains the maximum number of miners. 
As shown in Fig.~\ref{BlockTime}, the average block generating time in each network is around $ 300 $s, which just matches the target average BGR, and indicates that the difficulty re-targeting scheme works well. 
In addition, comparing Type I, II, III and IV network, the fluctuation range of the block generating time is gradually reduced. 
This fluctuation comes from the randomness of the process in finding the target hash value of the E-PoW block header. 
This is because that when the miners take more computing tasks, the blockchain system will be more stable in the time needed to generate a valid new block. 
The transaction delay can consequently be more stable and the number of extreme long block time can be efficiently reduced.


\subsubsection{Efficiency of the miners' computing power}
\begin{figure}
	\centerline{\includegraphics[width=0.45\textwidth]{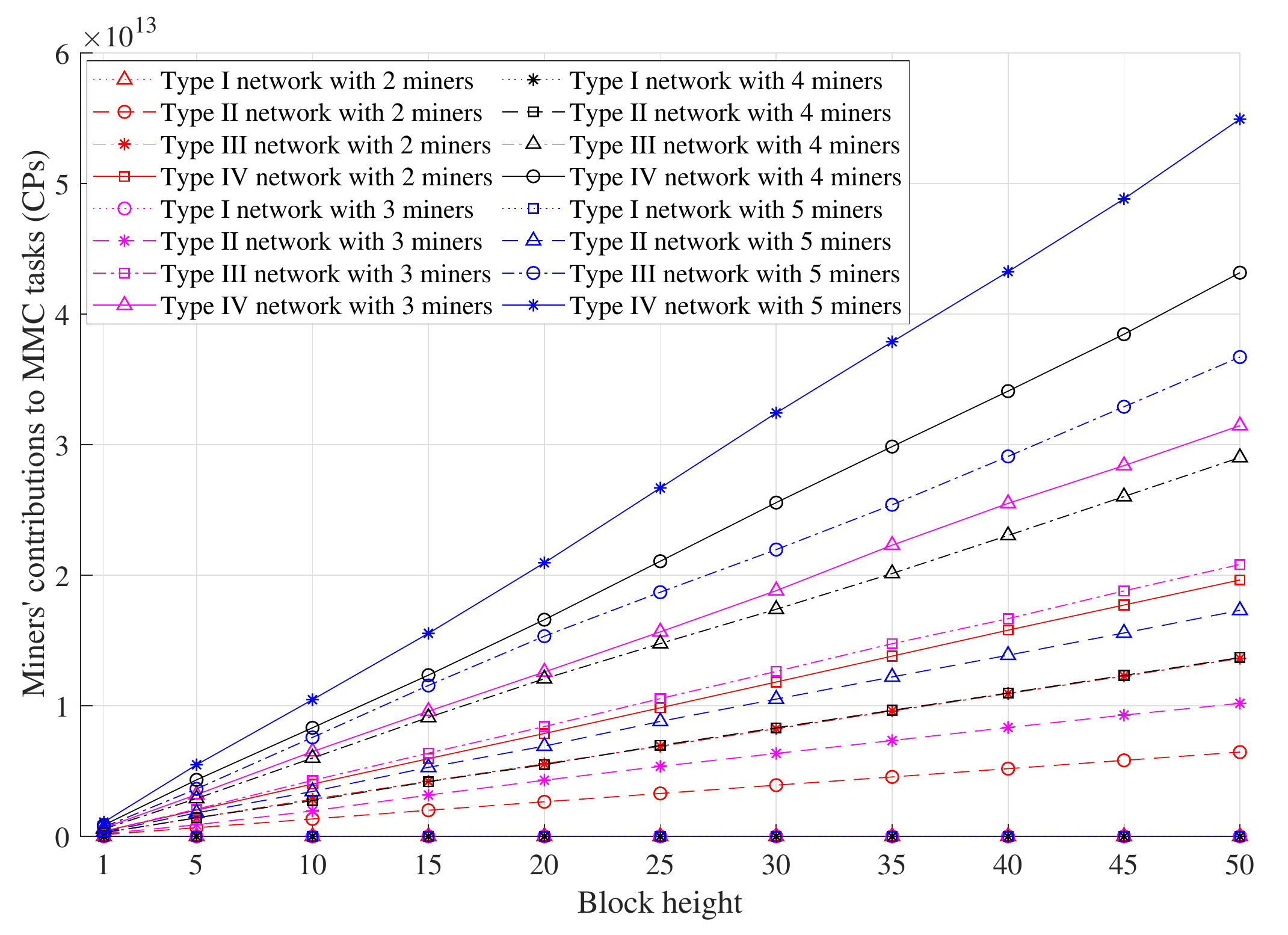}}
	\caption{The block height versus the miners' contributions to the MMC tasks.}
	\label{MatrixCalculation}
\end{figure}

Fig.~\ref{MatrixCalculation} illustrates the impact of the blockchain's height on the miners' contributions in computing MMC tasks. Specifically, in order to show the impact of the network scale, Fig.~\ref{MatrixCalculation} provides the trend of the miners' contributions when the number of miners varies from $2$ to $5$ in each type of network.

As shown in Fig.~\ref{MatrixCalculation},  the contributions of the miners to AI training are evaluated in the clock periods (CPs) they devote in computing MMC tasks.
In Type II, III and IV network, the miners' contributions to MMC tasks increase when the block height gets larger. That is to say, the amount of salvaged computing power in the miners will be continuously increasing with the operation of the blockchain.
Compared with Type I network, during the same time span, Type II, III, IV network not only generate $50$ valid blocks, but also provide MMC services for user devices. 
Specifically, among the four types of blockchain networks, Type IV network salvages the most amount of computing power for MMC tasks, followed by Type III network, Type II network, and Type I network in sequence. 
The reason is that the types of miners in these networks are different. The miners in Type IV network are most active in conducting task-involved loops, which means that they complete the most tasks in each block time. 
Thus, compared with other networks, the amount of effectively utilized computing power in Type IV network is the largest. 
Therefore, the computing power of E-PoW based network can be utilized more effectively than that of the traditional PoW-based network. The more active the miners are involved in the MMC tasks, the more computing power is effectively salvaged.

In each type of network, we can find the amount of the salvaged computing power for MMC tasks also increases with the expansion of the network scale. Take Type IV network as an example, when the network has only $ 2 $ miners and the block height is $50$, the miners' total computing resource salvaged for MMC tasks is about $ 2 \times 10^{13} $ CPs. 
When the number of miners grows to $5$, the salvaged computing resource increases to $ 5.5 \times 10^{13} $ CPs.
Consequently, unlike traditional PoW-based blockchains, the expansion of network scale will lead to more computing resource salvaged, or in another word, less computing resource wasted.

\begin{figure}
	\centerline{\includegraphics[width=0.48\textwidth]{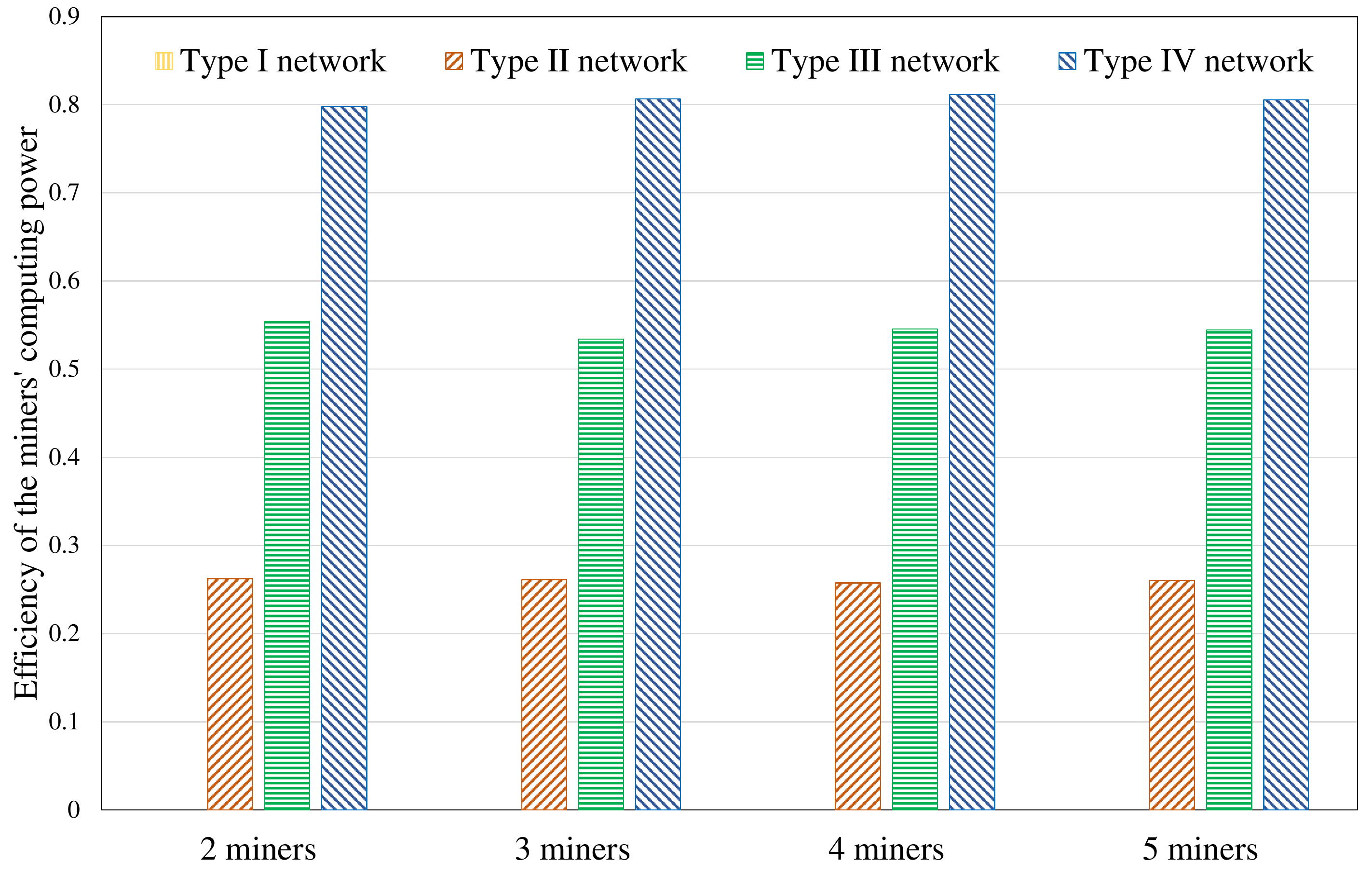}}
	\caption{Impact of network scale on the efficiency of the miners' computing power.}
	\label{UtilizationRate}
\end{figure}

Fig.~\ref{UtilizationRate} illustrates the impact of the network scale on the efficiency of the miners' computing power. 
For each network type, although the number of miners is different, the efficiency change of the computing power is negligible.
That is to say, the difference of the network scale has little influence on the efficiency of the computing power in the miners. At the same time, the efficiency of the computing power is proportional to the initiative of the miners in completing MMC tasks. In Type I network, which is a traditional PoW network, the miners utilize all their computing power for target hash value search. That is to say, besides the block mining, no computing power is salvaged and effectively used for MMC tasks.
As a comparison, in Type II, III and IV network, the number of nsub-tasks completed by each miner during one block time reaches their corresponding maximum value, and the efficiency of their computing power improves considerably by up to $ 80$ percent compared with that in PoW-based blockchains.

\section{Conclusion}
This paper has proposed an E-PoW consensus for 6G systems to integrate vast MMC tasks of AI training into the block mining of blockchains, as well as the detailed schemes transforming MMC tasks generated by AI to nsub-tasks computable in the miners. 
Based on the elaborated schemes in the consensus, E-PoW can keep the fairness, independence and security of generating valid blocks while connecting AI and blockchain in 6G systems.
The extensive experiments show that E-PoW can salvage by up to $80$ percent computing power from pure block mining for parallel AI training. 
For future work, a reward adjustment scheme can be designed to adjust the miners’ initiative of participating in the MMC tasks. To further improve the efficiency of the AI training, schemes to reduce the transmission delay of the tasks and results can also be studied.

\section*{Acknowledgment}
This work was partly supported by the National Key R\&D Program of China (No. 2018YFE0117500), the Science and Technology Program of Sichuan Province, China (No. 2019YFG0534), and the EU H2020 Project COSAFE (No. MSCA-RISE-2018-824019).

\bibliographystyle{IEEEtran}
\bibliography{connecting-AI-and-blockchain-in-6G}

\begin{thebibliography}{10}
\providecommand{\url}[1]{#1}
\csname url@samestyle\endcsname
\providecommand{\newblock}{\relax}
\providecommand{\bibinfo}[2]{#2}
\providecommand{\BIBentrySTDinterwordspacing}{\spaceskip=0pt\relax}
\providecommand{\BIBentryALTinterwordstretchfactor}{4}
\providecommand{\BIBentryALTinterwordspacing}{\spaceskip=\fontdimen2\font plus
\BIBentryALTinterwordstretchfactor\fontdimen3\font minus
  \fontdimen4\font\relax}
\providecommand{\BIBforeignlanguage}[2]{{%
\expandafter\ifx\csname l@#1\endcsname\relax
\typeout{** WARNING: IEEEtran.bst: No hyphenation pattern has been}%
\typeout{** loaded for the language `#1'. Using the pattern for}%
\typeout{** the default language instead.}%
\else
\language=\csname l@#1\endcsname
\fi
#2}}
\providecommand{\BIBdecl}{\relax}
\BIBdecl

\bibitem{CM_2019_AI_empowered_6G}
K.~B. {Letaief}, W.~{Chen}, Y.~{Shi}, J.~{Zhang}, and Y.~A. {Zhang}, ``The
  roadmap to 6g: Ai empowered wireless networks,'' \emph{IEEE Communications
  Magazine}, vol.~57, no.~8, pp. 84--90, 2019.

\bibitem{TNSE_2020_BATS}
R.~{Gupta}, A.~{Shukla}, and S.~{Tanwar}, ``Bats: A blockchain and ai-empowered
  drone-assisted telesurgery system towards 6g,'' \emph{IEEE Transactions on
  Network Science and Engineering}, pp. 1--1, 2020.

\bibitem{bitcoin}
S.Nakamoto, ``Bitcoin: A peer-to-peer electronic cash system,'' [Online].
  Available: \url{https://bitcoin.org/bitcoin.pdf}, 2008.

\bibitem{coincap}
``Top 100 cryptocurrencies by market capitalization,'' [Online]. Avaliable:
  \url{https://coinmarketcap.com}, 2021.

\bibitem{ppcoin}
S.~King and S.~Nadal, ``Ppcoin: peer-to-peer crypto-currency with
  proof-of-stake,'' [Online]. Available:
  \url{http://peercoin.net/assets/paper/peercoin-paper.pdf}, Aug. 2012.

\bibitem{poa}
I.~Bentov, C.~Lee, A.~Mizrahi, and M.~Rosenfeld, ``Proof of activity: Extending
  bitcoin's proof of work via proof of stake.'' \emph{IACR Cryptology ePrint
  Archive}, 2014.

\bibitem{dpos}
D.~Larimer, ``Delegated proof-of-stake (dpos),'' [Online]. Available:
  \url{https://bitshares.org/technology/delegated-proof-of-stake-consensus},
  2014.

\bibitem{PoL}
M.~Milutinovic, W.~He, H.~Wu, and M.~Kanwal, ``Proof of luck: An efficient
  blockchain consensus protocol,'' in \emph{proceedings of the 1st Workshop on
  System Software for Trusted Execution}, 2016, pp. 1--6.

\bibitem{Wei_2020_Profit_game}
Y.~{Wei}, M.~{Xiao}, N.~{Yang}, and S.~{Leng}, ``Block mining or service
  providing: A profit optimizing game of the pow-based miners,'' \emph{IEEE
  Access}, vol.~8, pp. 134\,800--134\,816, 2020.

\bibitem{pox}
A.~Shoker, ``Sustainable blockchain through proof of exercise,'' in \emph{2017
  IEEE 16th International Symposium on Network Computing and Applications
  (NCA)}.\hskip 1em plus 0.5em minus 0.4em\relax IEEE, 2017, pp. 1--9.

\bibitem{consensus_survey}
W.~{Wang}, D.~T. {Hoang}, P.~{Hu}, Z.~{Xiong}, D.~{Niyato}, P.~{Wang},
  Y.~{Wen}, and D.~I. {Kim}, ``A survey on consensus mechanisms and mining
  strategy management in blockchain networks,'' \emph{IEEE Access}, vol.~7, pp.
  22\,328--22\,370, 2019.

\bibitem{primecoin}
S.~King, ``Primecoin: Cryptocurrency with prime number proof-of-work,''
  [Online]. Available: \url{http://primecoin.io/bin/primecoin-paper.pdf}, 2013.

\bibitem{podl}
C.~{Chenli}, B.~{Li}, Y.~{Shi}, and T.~{Jung}, ``Energy-recycling blockchain
  with proof-of-deep-learning,'' in \emph{2019 IEEE International Conference on
  Blockchain and Cryptocurrency (ICBC)}, May 2019, pp. 19--23.

\bibitem{google}
N.~P. {Jouppi}, C.~{Young}, N.~{Patil} \emph{et~al.}, ``In-datacenter
  performance analysis of a tensor processing unit,'' in \emph{2017 ACM/IEEE
  44th Annual International Symposium on Computer Architecture (ISCA)}, June
  2017, pp. 1--12.

\end{thebibliography}

\begin{IEEEbiographynophoto}{Yunkai Wei}
(ykwei@uestc.edu.cn) is currently an Associate Professor with the School of Information and Communication Engineering, University of Electronic Science and Technology of China, where he received his B. Eng., M. Eng. and PhD degree. He was also a visiting researcher at California Institute of Technology during 2013-2014. His research interests include blockchain and machine learning, wireless communications and networks, and the Internet of Things.
\end{IEEEbiographynophoto}

\begin{IEEEbiographynophoto}{Zixian An}
(zxan@std.uestc.edu.cn) is a graduate student at the School of Information and Communication Engineering, University of Electronic Science and Technology of China. His research interests include blockchain and wireless networks.
\end{IEEEbiographynophoto}

\begin{IEEEbiographynophoto}{Supeng Leng}
(spleng@uestc.edu.cn) is a Full Professor in the School of Information \& Communication Engineering, University of Electronic Science and Technology of China (UESTC). He received his Ph.D. degree from Nanyang Technological University (NTU), Singapore. His research focuses on wireless communications and networks, blockchain and network security, and the Internet of Things. 
\end{IEEEbiographynophoto}

\begin{IEEEbiographynophoto}{Kun Yang}
[M'00, SM'07] (kunyang@essex.ac.uk) received his Ph.D. from the Department of Electronic \& Electrical Engineering, University College London, UK. He is currently a Chair Professor in the School of Computer Science \& Electronic Engineering, University of Essex, leading the Network Convergence Laboratory, UK. His research interests include blockchain and wireless networks, future Internet technology and network virtualization, mobile cloud computing and networking.
\end{IEEEbiographynophoto}

\end{document}